\documentclass[12pt,letterpaper]{article}
\usepackage{amssymb,amsfonts,amsthm,amsmath}
\usepackage{graphicx}
\usepackage{placeins}
\usepackage{natbib}

\def\hang{\hangindent\parindent}
\def\rf{\par\noindent\hang}
\def\ssk{\smallskip}

\def\rf{\par\noindent\hang}
\def\nin{\noindent}

\begin{document}

\baselineskip=24pt

\begin{center} {\Large{\textbf {Effect of a preliminary test of homogeneity of stratum-specific
odds ratios on their confidence intervals}}}
\end{center}

\medskip

\smallskip


\begin{center}
\large{Paul Kabaila$^*$  and Dilshani Tissera}
\end{center}

\begin{center}
{\sl Department of Mathematics and Statistics, La Trobe University, Victoria 3086, Australia}.
\end{center}

\vspace{11cm}

\nin $^*$ Author to whom correspondence should be addressed.
Department of Mathematics and Statistics, La Trobe University,
Victoria 3086, Australia.
Tel.: +61 3 9479 2594, Fax:
 +61 3 9479 2466,
 {E-mail:} P.Kabaila@latrobe.edu.au

\newpage

\begin{center}
{\large Abstract}
\end{center}

\ssk

\noindent Consider a case-control study in which the aim is to assess a factor's effect
 on disease occurrence. We suppose that this factor is dichotomous. Also suppose that the data consists of two strata, each stratum summarized by a two-by-two table. A commonly-proposed two-stage analysis of this type of data is the following. We carry out a preliminary test of homogeneity of the stratum-specific odds ratios. If the null hypothesis of homogeneity is accepted then we find a confidence interval for the assumed common value (across strata) of the odds ratio. We examine the statistical
properties of this  two-stage analysis, based on the Woolf method, on confidence intervals constructed for the stratum-specific odds ratios, for large numbers of cases and controls for each stratum. We provide
both a Monte Carlo simulation method and an elegant large-sample method for this examination. These methods are applied to obtain numerical
results in the context of the large numbers of cases and controls for each stratum that arose in a real-life dataset.
In this context, we find that the preliminary test of homogeneity of the stratum-specific odds ratios has a very harmful effect on the coverage probabilities of these confidence intervals.

\bigskip

\bigskip


\newpage

\noindent{\large {\bf 1. Introduction}}

\medskip

\noindent Consider a case-control study in which the aim is to assess a factor's
effect on disease occurrence. We suppose that this factor is
dichotomous. Also suppose that the data consists of two strata, each stratum summarized by a
$2 \times 2$ table. The parameters of interest are the
stratum-specific odds ratios.
A commonly-proposed two-stage analysis of this type of data is the following, see e.g.
section 4.4 of Breslow and Day (1980), Section 16.2 of Pagano and Gauvreau (2000)
and Section 13.6 of Rosner (2011).
 We carry out a preliminary test of homogeneity of the stratum-specific
odds ratios. If the null hypothesis of homogeneity is accepted then
we find a confidence interval for the assumed common value (across strata) of the odds ratio.

From a practical point of view, we must state what action we take when the
null hypothesis of homogeneity of stratum-specific odds ratios is rejected. It would not
make sense for a consulting statistician to tell a client that this null hypothesis
has been rejected and so the statistician will do nothing. There is some awareness of
the need to clearly state what action we take when the null hypothesis of homogeneity
is rejected, see e.g. p.279 of Rothman et al (2008) and p.620 of Rosner (2011). The latter
states that ``If the true {\sl OR}s are significantly different, then it makes no
sense to obtain a pooled-{\sl OR} estimate ... Instead, separate {\sl OR}s
should be reported''. We suppose that when the null
hypothesis of homogeneity is rejected, we compute confidence intervals
for each of the stratum-specific odds ratios.

Our aim is to examine the statistical properties of this two-stage analysis, in the context of
simultaneous inference for the stratum-specific odds ratios. In Section 2, we
provide a precise general formulation of this two-stage analysis.
We examine the statistical properties of this two-stage analysis
using the Woolf method (described e.g. on p.139 of Breslow and Day, 1980)
to carry out the preliminary test of homogeneity of the stratum-specific odds ratios and to
construct confidence intervals for the stratum-specific odds ratios,
for two strata and large numbers of cases and controls for each stratum. We provide
both a Monte Carlo simulation method and an elegant large-sample method for this examination. These
methods are applied to obtain numerical
results in the context of
case and control sample sizes that come from a study whose aim is to assess the effect of the
consumption of caffeinated coffee on nonfatal myocardial infarctions for adult males under the age of 55
(Pagano and Gauvreau, 2000 and Rosenberg et al, 1988).
Our general conclusion is that
the preliminary test of homogeneity of the stratum-specific odds ratios
has a very harmful effect on the coverage probabilities of the confidence intervals
for these odds ratios,
for two strata when the numbers of cases and controls in each
stratum is large.

\bigskip

\noindent {\large{ \textbf{2. Precise general formulation of the two-stage analysis}}}

\medskip

\noindent For easier cross-referencing with the notation used in Section 3,
we phrase our discussion in terms of log odds ratios.
Let $\theta_i$ denote the log odds ratio for the $i$ th stratum ($i=1,2$). To provide
a precise general formulation of the two-stage analysis, our first step is to
describe what we would do if it was known with certainty (a) that $\theta_1 \ne \theta_2$
and (b) that $\theta_1 = \theta_2$. We consider simultaneous inference for $\theta_1$
and $\theta_2$. Consequently, this description is in terms of simultaneous confidence
intervals for $\theta_1$ and $\theta_2$.

\noindent $\bullet$ \textbf{Suppose that} $\boldsymbol{\theta_1 \ne \theta_2}$

\noindent Use the confidence interval $\hat{I}_1$ for $\theta_1$ based solely on the two-by-two
table for stratum 1, with approximate coverage $\sqrt{1-\alpha}$.
Also, use the confidence interval $\hat{I}_2$ for $\theta_2$ based solely on the two-by-two
table for stratum 2, with approximate coverage $\sqrt{1-\alpha}$.
The confidence intervals $\hat{I}_1$ and $\hat{I}_2$ have simultaneous coverage
approximately $1-\alpha$, since
$P \big(\theta_1 \in \hat{I}_1, \theta_2 \in \hat{I}_2 \big)
= P \big(\theta_1 \in \hat{I}_1 \big) \, P \big(\theta_2 \in \hat{I}_2 \big) \approx 1-\alpha$.

\noindent $\bullet$ \textbf{Suppose that} $\boldsymbol{\theta_1 = \theta_2}$

\noindent Let $\hat{J}$ be the confidence interval for $\theta=\theta_1=\theta_2$ based
on the two-by-two
tables for both strata, with approximate coverage $1-\alpha$.
Let
$\hat{J}_1 = \hat{J}$ and $\hat{J}_2 = \hat{J}$
be confidence intervals for $\theta_1$ and $\theta_2$, respectively. These confidence
intervals have simultaneous coverage approximately $1-\alpha$, since
$P \big(\theta_1 \in \hat{J}_1, \theta_2 \in \hat{J}_2 \big)
= P \big(\theta \in \hat{J} \big)  \approx 1-\alpha$.

The two-stage analysis is precisely formulated as follows. If the null hypothesis of homogeneity is
rejected then we use the confidence intervals $\hat{I}_1$ and $\hat{I}_2$ for $\theta_1$ and
$\theta_2$, respectively. If, on the other hand, this null hypothesis is
accepted then we use the confidence intervals $\hat{J}_1$ and $\hat{J}_2$ for $\theta_1$ and
$\theta_2$, respectively. The nominal simultaneous coverage of the resulting confidence
intervals for $\theta_1$ and $\theta_2$ is $1-\alpha$. We will assess this two-stage analysis
by comparing the actual simultaneous coverage probability of these confidence intervals with
$1-\alpha$.
Of course, there are several different possible choices of preliminary test of homogeneity
and confidence intervals that can be used in the two-stage analysis. As explained
in the next section, we use tests and confidence intervals based on the Woolf method.

\bigskip

\noindent {\large{ \textbf{3. The two-stage analysis that will be evaluated}}}

\medskip

\noindent We use the following notation for the $2 \times 2$ contingency table
that summarizes the data for the $i$ th stratum ($i=1, 2$). Let $n_i$ denote
the number of subjects with the disease (cases), with $y_i$ of these subjects
exposed to the factor. Also, let $n_i^{\prime}$ denote
the number of subjects without the disease (controls), with $y_i^{\prime}$ of these subjects
exposed to the factor. We use upper case to denote random variables and lower case to
denote observed values. Thus, for example, $Y_i$ is the random variable corresponding
to the observed value $y_i$. We use the following model for the data in this table.
The random variables $Y_i$ and $Y_i^{\prime}$ are independent, with
$Y_i \sim \text{Binomial}(n_i, p_i)$ and $Y_i^{\prime} \sim \text{Binomial}(n_i^{\prime}, p_i^{\prime})$.
Let $\epsilon$ be a specified small positive number ($0 < \epsilon < \frac{1}{2}$).
Suppose that $p_i \in [\epsilon, 1-\epsilon]$ and $p_i^{\prime} \in [\epsilon, 1-\epsilon]$,
for $i=1, 2$.
The parameter of interest for this table is the odds ratio
\begin{equation*}
\psi_i = \frac{p_i/(1-p_i)}{p_i^{\prime}/(1-p_i^{\prime})}.
\end{equation*}
We find a confidence interval for $\psi_i$ as follows.
We first find a confidence interval for the log odds ratio $\theta_i = \ln(\psi_i)$
and then transform this in the obvious way into a confidence interval
for the odds ratio $\psi_i$.

We consider the two-stage analysis, described in Section 2,
implemented using Woolf's method. The maximum likelihood estimates of
$p_i$ and $p_i^{\prime}$ are $\tilde{p}_i=y_i/n_i$ and
$\tilde{p}_i^{\prime}=y_i^{\prime}/n_i^{\prime}$, respectively.
The resulting estimator of $\theta_i$ is
\begin{equation*}
\tilde{\Theta}_i
= \ln \left ( \frac{\tilde{p}_i/(1-\tilde{p}_i)}{\tilde{p}_i^{\prime}/(1-\tilde{p}_i^{\prime})} \right ).
\end{equation*}
This estimator has a number of disadvantages, including the fact that it is undefined for
$y_i$ either 0 or $n_i$ and for $y_i^{\prime}$ either 0 or $n_i^{\prime}$. We do not use
this estimator. Instead, we follow the common recommendation (see e.g. page 139 of Breslow and Day, 1980)
of estimating $p_i$ and $p_i^{\prime}$ by
$\hat{p}_i = (y_i + \frac{1}{2})/(n_i + 1)$ and
$\hat{p}_i^{\prime} = (y_i^{\prime} + \frac{1}{2})/(n_i^{\prime} + 1)$, respectively.
The resulting estimator of $\theta_i$ is
\begin{equation*}
\hat{\Theta}_i
= \ln \left ( \frac{\hat{p}_i/(1-\hat{p}_i)}{\hat{p}_i^{\prime}/(1-\hat{p}_i^{\prime})} \right ).
\end{equation*}
This estimator has the following three advantages. Firstly, it is defined for all
possible values of $y_i$ and $y_i^{\prime}$. Secondly, according to page 32 of Cox and Snell (1989),
$\hat{\Theta}_i$ is an asymptotically less biased estimator of $\theta_i$ than $\tilde{\Theta}_i$.
Thirdly, the use of this type of adjustment of the maximum likelihood estimates
of $p_i$ and $p_i^{\prime}$ can be remarkably
effective in improving the coverage probability properties of Wald-type
confidence intervals based on these estimates, see e.g.
Agresti and Caffo (2000).

 Woolf's method is based on the
approximation that
\begin{equation*}
\hat{\Theta}_i \sim N(\theta_i, \sigma_i^2),
\end{equation*}
where
\begin{equation*}
\sigma_i^2 = \frac{1}{n_i} \left ( \frac{1}{p_i} + \frac{1}{1-p_i} \right)
+ \frac{1}{n_i^{\prime}} \left ( \frac{1}{p_i^{\prime}} + \frac{1}{1-p_i^{\prime}} \right),
\end{equation*}
and the approximation that $\sigma_i^2$ is equal to
\begin{equation*}
\hat{\sigma}_i^2 = \frac{1}{n_i} \left ( \frac{1}{\hat{p}_i} + \frac{1}{1-\hat{p}_i} \right)
+ \frac{1}{n_i^{\prime}} \left ( \frac{1}{\hat{p}_i^{\prime}} + \frac{1}{1-\hat{p}_i^{\prime}} \right).
\end{equation*}

We test
the null hypothesis of homogeneity $H_0: \theta_1 = \theta_2$ against the alternative hypothesis $H_1$ that
the $\theta_1 \ne \theta_2$.
We carry out this test using the test statistic
\begin{equation*}
\hat{T} = \frac{\hat{\Theta}_1 - \hat{\Theta}_2}{\sqrt{\hat{\sigma}_1^2 + \hat{\sigma}_2^2}}.
\end{equation*}
We make the approximation that $\hat{T} \sim N(0,1)$ under $H_0$. Let $\beta$ denote the nominal level of
significance of this test.

If $H_0$ is rejected then the
confidence intervals for $\theta_1$ and $\theta_2$, with nominal simultaneous
coverage $1-\alpha$, are $\hat{I}_1$ and $\hat{I}_2$ respectively,
where
\begin{equation*}
\hat{I}_i = \big [ \hat{\Theta}_i - \tilde{c}_{\alpha} \, \hat{\sigma}_i, \, \hat{\Theta}_i + \tilde{c}_{\alpha} \, \hat{\sigma}_i \big]
\end{equation*}
with $\tilde{c}_{\alpha}$ defined by $P \big( -\tilde{c}_{\alpha} \le Z \le \tilde{c}_{\alpha} \big) = \sqrt{1 - \alpha}$ for $Z \sim N(0,1)$.
If, on the other hand, $H_0$ is accepted then we
carry out inference based on the assumption that
$\theta_1 = \theta_2 = \theta$. Define
\begin{equation*}
\hat{\Theta} = \frac{(\hat{\Theta}_1/\hat{\sigma}_1^2)+(\hat{\Theta}_2/\hat{\sigma}_2^2)}
{(1/\hat{\sigma}_1^2)+(1/\hat{\sigma}_2^2)},
\end{equation*}
which is the estimator of $\theta$ assuming that $\theta_1 = \theta_2 = \theta$.
If $\theta = \theta_1 = \theta_2$ then the following confidence interval for $\theta$ has nominal coverage $1-\alpha$:
\begin{equation*}
\hat{J} = \left [ \hat{\Theta} - c_{\alpha} \left(  \frac{1}{\hat{\sigma}_1^2}+ \frac{1}{\hat{\sigma}_2^2}\right )^{-1/2}, \,
\hat{\Theta} + c_{\alpha} \left(  \frac{1}{\hat{\sigma}_1^2}+ \frac{1}{\hat{\sigma}_2^2}\right )^{-1/2} \right ].
\end{equation*}
where $c_{\alpha}$ is defined by $P \big( -c_{\alpha} \le Z \le c_{\alpha} \big) = 1 - \alpha$ for $Z \sim N(0,1)$.
Let $\hat{J}_1=\hat{J}$ and $\hat{J}_2=\hat{J}$. If $H_0$ is accepted then
the confidence intervals for $\theta_1$ and $\theta_2$, with nominal simultaneous
coverage $1-\alpha$, are $\hat{J}_1$ and $\hat{J}_2$,
respectively.

\bigskip

\noindent {\large{ \textbf{4. Application to case and control sample sizes that
arise in a real-life data set}}}

\medskip

\noindent Consider the case-control study data with two  strata, described on p. 376 of Pagano and Gauvreau (2000). This data
originates from Rosenberg et al (1988). Pagano and Gauvreau (2000) carry out a preliminary test of homogeneity of the odds
ratios for these 2 strata, which is almost identical to that described in the previous section. They
conclude that they cannot reject the null hypothesis of the odds ratios being the same for these 2 strata.
They then use the Mantel-Haenszel method to estimate the odds ratio, which is assumed to be the same
for both of these strata. However, the Mantel-Haenszel method is inefficient, in the context of a
fixed number of strata and large numbers of cases and controls for each stratum, unless special
circumstances hold (Tarone et al, 1983). This is one of the reasons why we estimate the common odds ratio from the two
strata in the previous section using Woolf's method. The other reason for doing this is that this
permits us to find the elegant large-sample approximation described in Section 5.

For the data described on p.376 of Pagano and Gauvreau (2000), $k=2$, $n_1 = 1092$, $n_1^{\prime} = 467$, $n_2 = 449$ and
$n_2^{\prime} = 488$. The parameters of interest are the stratum-specific log odds ratios
\begin{equation*}
\theta_1 = \ln \left ( \frac{p_1/(1-p_1)}{p_1^{\prime}/(1-p_1^{\prime})} \right ) \ \ \
\text{and} \ \ \
\theta_2 = \ln \left ( \frac{p_2/(1-p_2)}{p_2^{\prime}/(1-p_2^{\prime})} \right ).
\end{equation*}
Our aim is to find confidence intervals for $\theta_1$ and $\theta_2$ with simultaneous
coverage $1-\alpha$. We suppose that $1-\alpha = 0.95$.
We also suppose that $(p_1, p_1^{\prime}, p_2, p_2^{\prime})$ belongs to the set
$A = [0.02, 0.98]^4$.
Under this restriction, it is expected that the distributions of $\hat{\Theta}_1$ and $\hat{\Theta}_2$ will be
close to normal. To see this, consider the rule-of-thumb that the
cdf of $Y \sim \text{Binomial}(n,p)$ is approximated well by the $N(np, np(1-p))$ cdf
if $np(1-p) \ge 5$ (see e.g. p.133 of Rosner, 2011). Note that $n_1 p_1 (1-p_1) \ge 21.403$,
$n_1^{\prime} p_1^{\prime} (1-p_1^{\prime}) \ge 9.153$, $n_2 p_2 (1-p_2) \ge 8.800$ and
$n_2^{\prime} p_2^{\prime} (1-p_2^{\prime}) \ge 9.565$ for all $(p_1, p_1^{\prime}, p_2, p_2^{\prime})$ in
$A = [0.02, 0.98]^4$. Thus, the distributions of $Y_1$, $Y_1^{\prime}$, $Y_2$ and $Y_2^{\prime}$ will be
close to normal for all $(p_1, p_1^{\prime}, p_2, p_2^{\prime})$ in
$A = [0.02, 0.98]^4$. Consequently, we expect the distributions of $\hat{\Theta}_1$ and $\hat{\Theta}_2$ to be
close to normal.


Firstly, consider the Woolf method confidence intervals $\hat{I}_1$ and $\hat{I}_2$,
when we do {\sl not} carry out a preliminary test of homogeneity of the stratum-specific
odds ratios. Because $n_1$, $n_1^{\prime}$, $n_2$ and
$n_2^{\prime}$ are large, we expect that the simultaneous coverage probability
$P(\theta_1 \in \hat{I}_1, \theta_2 \in \hat{I}_2)$ will not fall far below $1-\alpha=0.95$
for all $(p_1, p_1^{\prime}, p_2, p_2^{\prime})$ in $A$.
Using the simulation method described in Appendix A, we obtained the rough estimate
0.951 of the minimum simultaneous coverage probability.
This coverage probability
is attained at $(p_1, p_1^{\prime}, p_2, p_2^{\prime})=(0.596, 0.788, 0.308, 0.788)$.
All of the computations presented in this paper were performed with programs written in MATLAB, using the
statistics toolbox.

Now consider the two-stage analysis.
Suppose that the nominal level of significance of the preliminary hypothesis test
is 0.05.
Using the simulation method described in Appendix A, we obtained the rough estimate
0.131 of the minimum simultaneous coverage probability.
This coverage probability
is attained at $(p_1, p_1^{\prime}, p_2, p_2^{\prime})=(0.692, 0.596, 0.02, 0.02)$.
Actually, this estimate is an accurate Monte Carlo simulation estimate of an {\sl upper bound} to
the minimum simultaneous coverage probability. This shows that the
confidence intervals resulting from the two-stage analysis
are completely inadequate.

\bigskip

\noindent {\large{ \textbf{5. The large-sample approximation}}}

\medskip

\noindent Note that $\hat{\Theta}_1$ and $\hat{\Theta}_2$ are independent random variables.
It may be proved that $(\hat{\Theta}_1-\theta_1)/\sigma_1$ and $(\hat{\Theta}_2-\theta_2)/\sigma_2$
both converge in distribution to $N(0,1)$
(as $\min(n_1, n_1^{\prime}, n_2, n_2^{\prime}) \rightarrow \infty$). It may also be
proved that $\hat{\sigma}_1$ and $\hat{\sigma}_2$ converge in probability to
$\sigma_1$ and $\sigma_2$, respectively (as $\min(n_1, n_1^{\prime}, n_2, n_2^{\prime}) \rightarrow \infty$).
So,
the large-sample approximation that we will use to analyze the procedure
described in the previous section is as follows. Firstly, $\hat{\Theta}_i$ has
an $N(\theta_i, \sigma_i^2)$ distribution, when both $n_i$ and
$n_i^{\prime}$ are large ($i=1,2$). Secondly,
in the expressions for $\hat{\Theta}$, $\hat{T}$, $\hat{J}$ and $\hat{I}_i$ (given in the previous section), we may replace $\hat{\sigma}_i$
by $\sigma_i$ ($i=1,2$).
Thirdly, we assume that $\sigma_1^2$ and $\sigma_2^2$ are known.
In Appendix C, we apply this approximation to obtain a large-sample
approximation to the simultaneous coverage probability of the confidence intervals for $\theta_1$ and
$\theta_2$, with nominal simultaneous coverage $1-\alpha$, resulting from the two-stage analysis.

\bigskip

\noindent {\large{ \textbf{6. Numerical results obtained using the large-sample \newline approximation}}}

\medskip

\noindent The case-control study data described in Section 3 consists of
two strata with sample sizes $n_1 = 1092$, $n_1^{\prime} = 467$, $n_2 = 449$ and
$n_2^{\prime} = 488$. In the present section, we consider the same number of strata and
the same sample sizes.
Our aim is to find confidence intervals for $\theta_1$ and $\theta_2$ with simultaneous
coverage $1-\alpha$.
As in Section 3, we suppose that
$(p_1, p_1^{\prime}, p_2, p_2^{\prime})$ belongs to the set
$A = [0.02, 0.98]^4$.

Firstly, consider the Woolf method confidence intervals $\hat{I}_1$ and $\hat{I}_2$,
when we do {\sl not} carry out a preliminary test of homogeneity of the stratum-specific
odds ratios. Obviously, the large-sample approximation described in Section 4
tells us that $P(\theta_1 \in \hat{I}_1, \theta_2 \in \hat{I}_2) = 1-\alpha$
for all $(p_1, p_1^{\prime}, p_2, p_2^{\prime})$ in $A$.

Now consider the two-stage analysis. As in Section 3, suppose that $1-\alpha = 0.95$.
Also suppose that the nominal level of significance of the preliminary hypothesis test
is 0.05. Define the step length $h = 0.096$.
The large-sample approximation to the coverage probability, described in detail in Appendix C, was
computed for each $(p_1, p_1^{\prime}, p_2, p_2^{\prime})$
belonging to the set $(0.02, 0.02+h, 0.02+2h, \ldots, 0.98)^4$.
The minimum value of this large-sample approximation was found to be 0.134847.
This value was achieved at $(p_1, p_1^{\prime}, p_2, p_2^{\prime})$ taking any one of the following values:
(0.692, 0.596, 0.02, 0.02), (0.308, 0.404, 0.02, 0.02), (0.692, 0.596, 0.98, 0.98)
and (0.308, 0.404, 0.98, 0.98). The large-sample approximation is a smooth function of
$(p_1, p_1^{\prime}, p_2, p_2^{\prime})$ and so this minimum value can be expected to be an accurate
approximation to large sample approximation minimized over $(p_1, p_1^{\prime}, p_2, p_2^{\prime})$ in $A$.

For the two-stage analysis, the simulation estimate of the minimum simultaneous coverage probability of
the confidence intervals for the stratum-specific odds ratios was found to be 0.131. This is quite close to
the minimum value of the large-sample approximation to the simultaneous coverage probability of
these confidence intervals, which was found to be 0.134847.
In this context, we find that the preliminary test of homogeneity of the stratum-specific odds
ratios has a very harmful effect on the coverage probabilities of these confidence intervals.

\bigskip

\noindent{\large{\textbf{7. The simultaneous coverage probability of the confidence intervals resulting from the two-stage analysis
is small away from the boundaries of the parameter space}}}

\medskip

\noindent In this section we deal exclusively with the confidence intervals resulting from the two-stage
analysis. As noted in the previous sections, the minimum simultaneous coverage probability is achieved
on the boundary of the parameter space $A$ for both the simulation and large-sample estimates of this
minimum coverage probability. If this simultaneous coverage
probability is small only at or near the boundaries of the parameter space then it might be argued that
statistical practitioners need not be concerned about the smallness of the minimum simultaneous coverage
probability. Therefore, it is natural to ask the question: Is this simultaneous coverage
probability small only at or near the boundaries of the parameter space?

In this section, we show that this simultaneous coverage probability is also small far from the
boundaries of the parameter space $A$. We do this as follows.
Suppose that $(p_1, p_1^{\prime}, p_2, p_2^{\prime})$ belongs to the set $A=[\epsilon, 1-\epsilon]^4$,
where $\epsilon$ is a small specified positive number ($0 < \epsilon < \frac{1}{2}$). Let
\begin{equation*}
\Delta = \sqrt{(1/n_1) + (1/n_2)} \qquad \text{and}
\qquad r = \frac{(1/n_1^{\prime})+(1/n_2^{\prime})}{(1/n_1)+(1/n_2)}
\end{equation*}
Also let $\Delta^{\prime} = r \, \Delta$. For each $p_1$ in $[\epsilon + \Delta, 1 - \epsilon - \Delta]$ and
$p_1^{\prime}$ in $[\epsilon + \Delta^{\prime}, 1 - \epsilon - \Delta^{\prime}]$, we find the minimum
over $p_2 = p_1 + \delta$ and $p_2^{\prime} = p_1^{\prime} - r \, \delta$, where $\delta$ is in
$[-\Delta, \Delta]$, of the large-sample simultaneous coverage probability of the
confidence intervals for $\theta_1$ and $\theta_2$ resulting from the two-stage analysis.
We then examine this partially-minimized coverage probability using a contour plot of it,
as a function of $(p_1, p_1^{\prime})$ in
$[\epsilon + \Delta, 1 - \epsilon - \Delta] \times [\epsilon + \Delta^{\prime}, 1 - \epsilon - \Delta^{\prime}]$.
Note that for $(p_1, p_1^{\prime})$ not close to the boundaries of this set, the partially-minimized
coverage is achieved at a value of $(p_1, p_1^{\prime}, p_2, p_2^{\prime})$ not close to the boundaries of $A$.
The large-sample analysis described in Appendix C includes the test statistic $T$ which has an
$N(\lambda,1)$ distribution, where $\lambda=(\theta_1 - \theta_2)/\sqrt{\sigma_1^2 + \sigma_2^2}$. In this
partial minimization, $\lambda$ is a function of $\delta \in [-\Delta, \Delta]$, for each given
$(p_1, p_1^{\prime})$. In Appendix D, we show that the range of this function includes the interval
$[-2,2]$. Therefore, this partial minimization includes the consideration of a wide interval of values
of $\lambda$, suggesting that the partially-minimized coverage will be quite small. Of course, whether or
not this is, indeed, the case needs to be assessed numerically.

Consider the case-control study data described in Section 3, which consists of
two strata with sample sizes $n_1 = 1092$, $n_1^{\prime} = 467$, $n_2 = 449$ and
$n_2^{\prime} = 488$. In this case, $\Delta = 0.056062$ and $r= 1.333316$,
so that $\Delta^{\prime} = 0.074748$. Figure 1
is a contour plot of the partially minimized coverage probability,
as a function of $(p_1, p_1^{\prime})$ in
$[0.02+\Delta, 0.98-\Delta] \times [0.02+\Delta^{\prime}, 0.98-\Delta^{\prime}]$, for $1-\alpha=0.95$. Figure 1 demonstrates
that the large-sample approximation to the simultaneous coverage probability is
much less than 0.95 for $(p_1, p_1^{\prime}, p_2, p_2^{\prime})$
far from the boundaries of the parameter space $A=[0.02,0.98]^4$.

It is straightforward to show that the harmful effect of the preliminary test of homogeneity of the stratum-specific odds ratios
does not disappear as the sample sizes increase. Consider
$n_1 = 1092 N$, $n_1^{\prime} = 467 N$, $n_2 = 449 N$ and $n_2^{\prime} = 488 N$, where $N$ is
a positive integer. It may be shown that, as we increase $N$, the partially-minimized coverage
probability converges to a limiting value for each $(p_1, p_1^{\prime})$. The contour
plot shown in Figure 1 does not differ greatly from the contour plot of this limiting value.
In other words, the harmful effect of the preliminary test of homogeneity of the stratum-specific odds ratios
does not disappear as the sample sizes increase.

\FloatBarrier

\begin{figure}[h]
\label{Figure1}
    \hspace{-1cm}
    \includegraphics[scale=0.78]{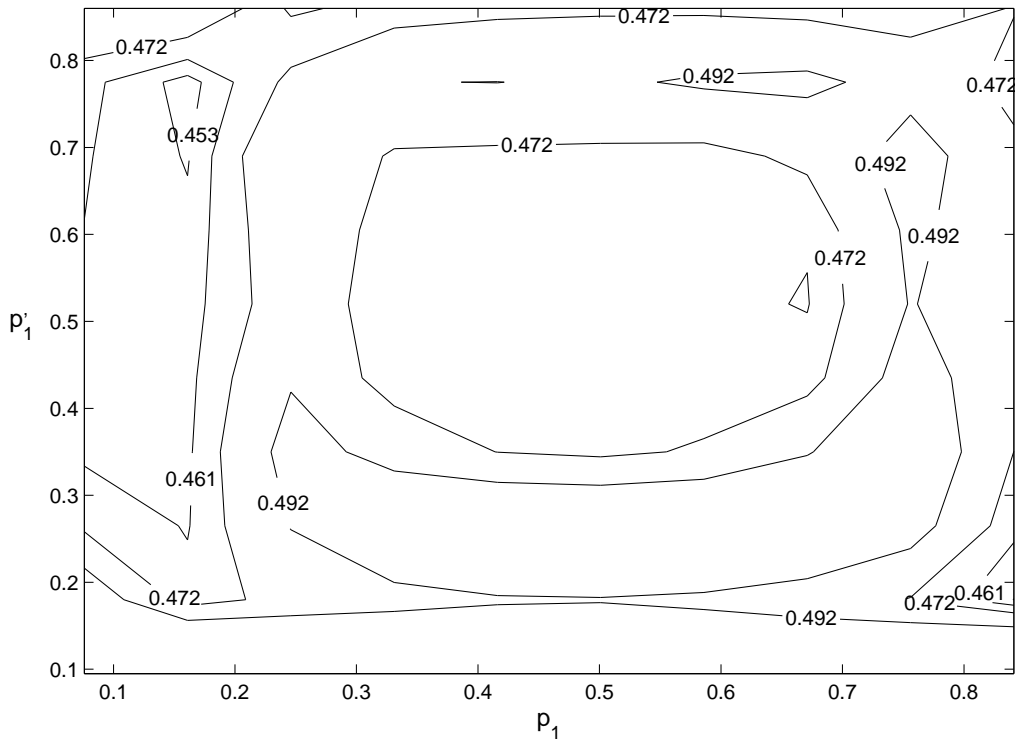}
    \caption{Plot of the partially-minimized coverage
probability of the confidence interval resulting from the two-stage analysis,
with nominal coverage probability 0.95 and nominal level of significance 0.05 of
the preliminary test of homogeneity. This plot shows the partially-minimized coverage
probability as a function of $(p_1, p_1^{\prime})$.}
\end{figure}
\FloatBarrier

\bigskip

\noindent{\large{\textbf{8. Discussion}}}

\medskip

\noindent The literature on the effect of preliminary statistical model selection (using, for example, hypothesis
tests or minimizing a criterion such as AIC) on confidence intervals
begins with the  work of Freeman (1989) who analyzed the effect of a preliminary test of the
null hypothesis of zero
differential carryover in a two-treatment two-period crossover trial on the
confidence interval for the difference of treatment effects.
This literature
has grown steadily since this work of Freeman and is reviewed by Kabaila (2009).
It is commonly the case that preliminary model selection has a highly detrimental effect on the
coverage probability of these confidence intervals.
However, each case (specified by a model, a model selection procedure and parameters of interest)
needs to be considered individually on its merits.

Our results show that the preliminary test of homogeneity of the stratum-specific odds ratios
should not be used. The harmful effect of this preliminary test is very substantial and exists far from
the boundaries of the parameter space. Furthermore, this harmful effect does not
disappear with increasing sample sizes.

\bigskip

\noindent{\large{\textbf{Acknowledgments}}}

\medskip

\noindent The authors are grateful to Ian Marschner for raising the question answered in Section 6 and
to Davide Farchione for his helpful comments.

\bigskip

\noindent{\large{\textbf{Appendix A: The search used to find an approximation to the minimum simultaneous coverage probability}}}

\medskip

\noindent In this appendix we describe the search through the parameter space that was used to find an approximation to the minimum
simultaneous coverage probability of specified confidence intervals for the log-odds ratios $\theta_1$ and $\theta_2$.
As shown in Appendix B for the particular case $P(\theta_1 \in \hat{I}_1, \theta_2 \in \hat{I}_2)$, this simultaneous coverage
probability is a discontinuous function of $(p_1, p_1^{\prime}, p_2, p_2^{\prime})$.
This makes it difficult to get a very accurate estimate of the simultaneous coverage
probability minimized over $(p_1, p_1^{\prime}, p_2, p_2^{\prime})$ in the parameter space $A$. Nonetheless,
the following search method provides a rough estimate of the minimum simultaneous coverage probability.

For a given value of $(p_1, p_1^{\prime}, p_2, p_2^{\prime})$, we estimate
the simultaneous coverage probability  of the confidence intervals by Monte Carlo simulation.
We use a search
method of the type described by Kabaila and Giri (2008) (cf. Section 3.1
of Kabaila and Leeb, 2006).
Define the step length $h = 0.096$. The simultaneous coverage probability of these confidence intervals
is estimated using $M = 10000$ simulations for each $(p_1, p_1^{\prime}, p_2, p_2^{\prime})$
belonging to the set $(0.02, 0.02+h, 0.02+2h, \ldots, 0.98)^4$. The 10 values of
$(p_1, p_1^{\prime}, p_2, p_2^{\prime})$ with the lowest estimates of this coverage probability
are then selected for further consideration. For each of these 10 values, the coverage probability
is then re-estimated using $M = 200000$ simulations. The value of $(p_1, p_1^{\prime}, p_2, p_2^{\prime})$
with the lowest estimate of this coverage probability is then selected for further consideration.
For this value, the coverage probability is then re-estimated using $M = 10^6$ simulations.

\bigskip

\noindent{\large{\textbf{Appendix B: Discontinuity of $\boldsymbol{P(\theta_1 \in \hat{I}_1, \theta_2 \in \hat{I}_2)}$
as a \newline function of $\boldsymbol{(p_1, p_1^{\prime}, p_2, p_2^{\prime})}$}}}

\medskip

\noindent Consider the simultaneous coverage probability
$P \big(\theta_1 \in \hat{I}_1, \theta_2 \in \hat{I}_2 \big)$.
Let $p = (p_1, p_1^{\prime}, p_2, p_2^{\prime})$ and $y = (y_1, y_1^{\prime}, y_2, y_2^{\prime})$.
Also let ${\cal Y}$ denote the set of possible values of $y$, so that
${\cal Y} = \{0, \ldots, n_1\} \times \{0, \ldots, n_1^{\prime}\} \times \{0, \ldots, n_2\} \times \{0, \ldots, n_2^{\prime}\}$.
Now let
$\hat{I}_i(y)$ denote the interval $\hat{I}_i$ evaluated at observed value $y$ ($i=1,2$).
Define
$B(p)$ to be the set of $y$ belonging to ${\cal Y}$ such that $\theta_1 \in \hat{I}_1(y)$ and $\theta_2 \in \hat{I}_2(y)$, for given $p$.
Note that
\begin{equation*}
P \big(\theta_1 \in \hat{I}_1, \theta_2 \in \hat{I}_2 \big) = \sum_{y \in B(p)} P(Y=y).
\end{equation*}
Whilst $P(Y=y)$ is a smooth function of $p$ for each $y \in {\cal Y}$, the set $B(p)$
changes as we change $p$. This leads to $P(\theta_1 \in \hat{I}_1, \theta_2 \in \hat{I}_2)$
being a discontinuous function of $p$.

\bigskip

\noindent{\large{\textbf{Appendix C: Details of the analysis using the large-sample \newline
approximation}}}

\medskip

\noindent Note that $\hat{\Theta}_1$ and $\hat{\Theta}_2$ are independent random variables.
The large-sample
approximation described in Section 4 is as follows.
The estimators
$\hat{\Theta}_1$ and $\hat{\Theta}_2$ have the following distributions:
$\hat{\Theta}_1 \sim N(\theta_1, \sigma_1^2)$ and $\hat{\Theta}_2 \sim N(\theta_2, \sigma_2^2)$,
where $\sigma_1^2$ and $\sigma_2^2$ are known. We test the null hypothesis of homogeneity
$H_0: \theta_1 = \theta_2$ against the alternative hypothesis $H_1: \theta_1 \ne \theta_2$, using
the test statistic
\begin{equation*}
T = \frac{\hat{\Theta}_1 - \hat{\Theta}_2}{\sqrt{\sigma_1^2 + \sigma_2^2}}.
\end{equation*}
Let $\beta$ denote the level of significance of this test. Note that $T \sim N(\lambda,1)$,
where $\lambda = (\theta_1-\theta_2)/\sqrt{\sigma_1^2 + \sigma_2^2}$. Define the quantile
$c_a$ by the requirement that $P(-c_a \le Z \le c_a) = 1 - a$ for $Z \sim N(0,1)$.
We accept $H_0$ if $|T| \le c_{\beta}$; otherwise we reject $H_0$.

If $H_0$ is rejected then the
confidence intervals for $\theta_1$ and $\theta_2$,
with nominal simultaneous
coverage $1-\alpha$, are $I_1$ and $I_2$ respectively,
where
\begin{equation*}
I_i = \big [ \hat{\Theta}_i - \tilde{c}_{\alpha} \, \sigma_i, \, \hat{\Theta}_i + \tilde{c}_{\alpha} \, \sigma_i \big]
\end{equation*}
with $\tilde{c}_{\alpha}$ defined by
$P \big( -\tilde{c}_{\alpha} \le Z \le \tilde{c}_{\alpha} \big) = \sqrt{1 - \alpha}$ for $Z \sim N(0,1)$.
Define
\begin{equation*}
\hat{\Theta} = \frac{(\hat{\Theta}_1/\sigma_1^2) + (\hat{\Theta}_2/\sigma_2^2)}{(1/\sigma_1^2) + (1/\sigma_2^2)},
\end{equation*}
which is the estimator of $\theta$, assuming that $\theta = \theta_1 = \theta_2$.
If $\theta = \theta_1 = \theta_2$ then the following confidence interval for $\theta$ has coverage $1-\alpha$:
\begin{equation*}
J = \left [ \hat{\Theta} - c_{\alpha} \left( \frac{1}{\sigma_1^2} + \frac{1}{\sigma_2^2}\right )^{-1/2}, \,
\hat{\Theta} + c_{\alpha} \left( \frac{1}{\sigma_1^2} + \frac{1}{\sigma_2^2}\right )^{-1/2} \right ].
\end{equation*}
Let $J_1=J$ and $J_2=J$. If $H_0$ is accepted then
the confidence intervals for $\theta_1$ and $\theta_2$, with nominal simultaneous
coverage $1-\alpha$, are $J_1$ and $J_2$,
respectively.

Our aim is to evaluate the coverage probability of the simultaneous confidence intervals
for $\theta_1$ and $\theta_2$ resulting from the above procedure for given
$n_i$, $n_i^{\prime}$, $p_i$ and $p_i^{\prime}$ ($i=1, 2$). By the law of total probability, this coverage probability
is equal to
\begin{equation*}
P \big(\theta_1 \in J, \theta_2 \in J, |T| \le c_{\beta} \big)
+ P \big(\theta_1 \in I_1, \theta_2 \in I_2, |T| > c_{\beta} \big).
\end{equation*}

To evaluate this coverage probability, we will make use of the following readily-established results.
The first result is that $\hat{\Theta}$ and $\hat{\Theta}_1 - \hat{\Theta}_2$
are independent random variables.
Since $\hat{\Theta}_1$ and $\hat{\Theta}_2$ are independent normally-distributed random variables,
$\big( \hat{\Theta}, \hat{\Theta}_1 - \hat{\Theta}_2 \big)$
has a bivariate normal distribution.
We therefore prove this result by showing that Cov$(\hat{\Theta}, \hat{\Theta}_1 - \hat{\Theta}_2) = 0$.
It is a corollary of this result that $\hat{\Theta}$ and $T$ are independent
random variables.
Now $\{ \theta_i \in J \} = \{ \hat{\Theta} \in K_i \}$, where
\begin{equation*}
K_i = \left [ \theta_i - c_{\alpha} \left( \frac{1}{\sigma_1^2} + \frac{1}{\sigma_2^2} \right)^{-1/2}, \,
\theta_i + c_{\alpha} \left( \frac{1}{\sigma_1^2} + \frac{1}{\sigma_2^2} \right)^{-1/2} \right ].
\end{equation*}
Thus
\begin{align*}
P \big(\theta_1 \in J, \theta_2 \in J, |T| \le c_{\beta} \big)
&= P \big(\hat{\Theta} \in K_1, \hat{\Theta} \in K_2, |T| \le c_{\beta} \big) \\
&= P \big(\hat{\Theta} \in K_1, \hat{\Theta} \in K_2 \big) \, P \big(|T| \le c_{\beta} \big) \\
&= P \big(\hat{\Theta} \in K_1 \cap K_2 \big) \, P \big(|T| \le c_{\beta} \big).
\end{align*}
Let $a$ denote the maximum of the lower endpoints of the intervals $K_1$ and $K_2$.
Also let $b$ denote the minimum of the upper endpoints of $K_1$ and $K_2$.
Observe that
\begin{equation*}
P( \hat{\Theta} \in K_1 \cap K_2) =
\begin{cases}
0 &\text{if } \ \ \ a \ge b \\
P(a \le \hat{\Theta} \le b) &\text{otherwise.}
\end{cases}
\end{equation*}
We compute $P(a \le \hat{\Theta} \le b)$ using the fact that $\hat{\Theta} \sim N(\theta_{av}, w)$,
where
\begin{equation*}
\theta_{av} = \frac{(\theta_1 / \sigma_1^2)+(\theta_2 / \sigma_2^2)}{(1 / \sigma_1^2)+(1/ \sigma_2^2)} \qquad \text{and}
\qquad w = \frac{1}{(1 / \sigma_1^2)+(1/ \sigma_2^2)}.
\end{equation*}

By the law of total probability,
\begin{equation*}
P(\theta_1 \in I_1, \theta_2 \in I_2) = P(\theta_1 \in I_1, \theta_2 \in I_2, |T| > c_{\beta})
+ P(\theta_1 \in I_1, \theta_2 \in I_2, |T| \le c_{\beta}).
\end{equation*}
Since $P(\theta_1 \in I_1, \theta_2 \in I_2) = 1 - \alpha$,
\begin{equation*}
P(\theta_1 \in I_1, \theta_2 \in I_2, |T| > c_{\beta}) =
1 - \alpha - P(\theta_1 \in I_1, \theta_2 \in I_2, |T| \le c_{\beta}).
\end{equation*}
We compute $P(\theta_1 \in I_1, \theta_2 \in I_2, |T| \le c_{\beta})$ using the following method.
Straightforward manipulations show that this probability is equal to
\begin{align*}
P \Bigg ( - \tilde{c}_{\alpha} \le &Z_1 \le \tilde{c}_{\alpha}, - \tilde{c}_{\alpha} \le Z_2 \le \tilde{c}_{\alpha}, \\
&\frac{\sigma_2}{\sigma_1} Z_2 - c_{\beta} \sqrt{1 + \frac{\sigma_2^2}{\sigma_1^2}} - \frac{\theta_1 - \theta_2}{\sigma_1}
\le Z_1 \le \frac{\sigma_2}{\sigma_1} Z_2 + c_{\beta} \sqrt{1 + \frac{\sigma_2^2}{\sigma_1^2}} - \frac{\theta_1 - \theta_2}{\sigma_1} \Bigg )
\end{align*}
where $Z_1 = (\hat{\Theta}_1 - \theta_1)/\sigma_1$ and $Z_2 = (\hat{\Theta}_2 - \theta_2)/\sigma_2$. Since
$Z_1$ and $Z_2$ are independent and identically $N(0,1)$ distributed, this probability is equal to
\begin{equation}
\label{integral}
\tag{C1}
\int_{-\tilde{c}_{\alpha}}^{\tilde{c}_{\alpha}} \int_{B(z_2)} \phi(z_1) \, dz_1 \, \phi(z_2) \, dz_2
\end{equation}
where $\phi$ denotes the $N(0,1)$ probability density function and
\begin{align*}
B(z_2) &= L_1 \cap L_2(z_2) \\
L_1 &= \big[-\tilde{c}_{\alpha}, \tilde{c}_{\alpha} \big] \\
L_2(z_2) &= \left [ \frac{\sigma_2}{\sigma_1} z_2 - c_{\beta} \sqrt{1 + \frac{\sigma_2^2}{\sigma_1^2}} - \frac{\theta_1 - \theta_2}{\sigma_1},
\frac{\sigma_2}{\sigma_1} z_2 + c_{\beta} \sqrt{1 + \frac{\sigma_2^2}{\sigma_1^2}} - \frac{\theta_1 - \theta_2}{\sigma_1} \right ].
\end{align*}
Let $\tilde{a}(z_2)$ denote the maximum of the lower endpoints of the intervals $L_1$ and $L_2(z_2)$.
Also let $\tilde{b}(z_2)$ denote the minimum of the upper endpoints of the intervals $L_1$ and $L_2(z_2)$. Let
\begin{equation*}
g(z_2) = \int_{B(z_2)} \phi(z_1) \, dz_1.
\end{equation*}
Observe that
\begin{equation*}
g(z_2) =
\begin{cases}
0 &\text{if } \ \ \ \tilde{a}(z_2) \ge \tilde{b}(z_2) \\
\Phi \big(\tilde{b}(z_2) \big) - \Phi \big(\tilde{a}(z_2) \big) &\text{otherwise}
\end{cases}
\end{equation*}
where $\Phi$ denotes the $N(0,1)$ distribution function.
Thus \eqref{integral} is equal to
\begin{equation}
\label{integral_simpler}
\tag{C2}
\int_{-\tilde{c}_{\alpha}}^{\tilde{c}_{\alpha}} g(z_2) \, \phi(z_2) \, dz_2
\end{equation}
Note that $g(z_2)$ is a very
smooth function of
$z_2 \in \big [-\tilde{c}_{\alpha}, \tilde{c}_{\alpha} \big]$,
except at a finite number (up to 4) values of $z_2$, where this function
is continuous but does not possess a first derivative. Therefore, \eqref{integral_simpler} is computed
by adding the numerical integrals over the obvious subintervals that have at least one of these
values of $z_2$ as an endpoint.

\bigskip

\noindent{\large{\textbf{Appendix D: A property of the partial minimization of \newline
the approximate coverage considered in Section 6}}}

\medskip

\noindent The large-sample analysis described in Appendix C includes the test statistic $T$ which has an
$N(\lambda,1)$ distribution, where $\lambda=(\theta_1 - \theta_2)/\sqrt{\sigma_1^2 + \sigma_2^2}$. In the
partial minimization described in the second paragraph of Section 6,
$\lambda$ is a function of $\delta \in [-\Delta, \Delta]$, for each given
$(p_1, p_1^{\prime})$. In this appendix, we show that the range of this function includes the interval
$[-2,2]$.

Suppose that $p_2 = p_1 + \delta$ and $p_2^{\prime} = p_1^{\prime} + \delta^{\prime}$, where
$|\delta|$ and $|\delta^{\prime}|$ are small. By Taylor expansion,
\begin{equation*}
\lambda \approx  \frac{\displaystyle{\frac{1}{p_1^{\prime}(1-p_1^{\prime})}} \delta^{\prime} - \frac{1}{p_1(1-p_1)} \delta}
{\sqrt{ \displaystyle{\left ( \frac{1}{n_1} + \frac{1}{n_2} \right ) \frac{1}{p_1(1-p_1)} +
\left ( \frac{1}{n_1^{\prime}} + \frac{1}{n_2^{\prime}} \right) \frac{1}{p_1(1-p_1)}}}}.
\end{equation*}
Now suppose that $\delta^{\prime} = - r \delta$, where $r$ is defined in Section 6. Thus
\begin{equation*}
\lambda \approx  \left (\frac{-1}{\displaystyle{\frac{1}{n_1} + \frac{1}{n_2}}} \right)
\sqrt{ \displaystyle{\left ( \frac{1}{n_1} + \frac{1}{n_2} \right ) \frac{1}{p_1(1-p_1)} +
\left ( \frac{1}{n_1^{\prime}} + \frac{1}{n_2^{\prime}} \right) \frac{1}{p_1(1-p_1)}}} \ \ \ \delta.
\end{equation*}
Since $p_1(1-p_1) \le 1/4$ and $p_1^{\prime}(1-p_1^{\prime}) \le 1/4$, $|\lambda| \ge 2 |\delta|/\Delta$,
where $\Delta$ is defined in Section 6.

\bigskip

\noindent{\large{\textbf{References}}}

\rf Agresti A, Caffo B. Simple and effective confidence intervals for
proportions and differences of proportions result from adding two successes
and two failures. {\sl American Statistician} 2000; {\bf 54}: 280--288.

\rf Breslow NE, Day NE. {\sl Statistical Methods in Cancer Research.
Volume 1 - The analysis of case-control studies}.
International Agency for Research on Cancer: Lyon, 1980.

\rf Cox DR, Snell EJ. {\sl Analysis of Binary Data}. 2nd edition.
Chapman and Hall: London, 1989.

\rf Freeman P. The performance of the two-stage analysis of two-treatment,
two-period crossover trials. {\sl Statistics in Medicine} 1989; {\bf 8}:1421--1432.

\rf Kabaila P. The coverage properties of confidence regions after model
selection. {\sl International Statistical Review} 2009; {\bf 77}:405--414.

\rf Kabaila P, Giri K. The coverage probability of confidence intervals in
$2^r$ factorial experiments after preliminary hypothesis testing. {\sl
Australian \& New Zealand Journal of Statistics} 2008; {\bf 50}:69--79.

\rf Kabaila P, Leeb H. On the large-sample minimal coverage probability of
confidence intervals after model selection. {\sl Journal of the American
Statistical Association} 2006; {\bf 101}:619--629.

\rf Pagano M, Gauvreau K. {\sl Principles of Biostatistics}, second edition.
Duxbury: Pacific Grove, CA, 2000.


\rf Rosenberg L, Palmer JR, Kelly JP, Kaufman DW, Shapiro S. Coffee-drinking
and nonfatal myocardial infarction in men under 55 years of age.
{\sl American Journal of Epidemiology} 1988; {\bf 128}:570--578.

\rf Rosner, B. {\sl Fundamentals of Biostatistics}, 7th edition.
Brooks/Cole, Boston, 2011.

\rf Rothman, K.J., Greenland, S. \& Lash, T.L. (2008) Modern Epidemiology, 3rd
edition. Lippincott Williams \& Wilkins, Philadelphia, 2008.

\rf Tarone RF, Gart JJ, Hauck WW. On the asymptotic inefficiency of certain
noniterative estimators of a common relative risk or odds ratio. {\sl Biometrika}
1983; {\bf 70}:519--522.

\end{document}